\documentclass[12pt,fleqn]{article}
\usepackage{amsmath,amssymb}
\usepackage{bm}
\usepackage[dvips]{graphicx}
\allowdisplaybreaks[3]

\begin{document}

\title{Determining the $\Theta^+$ quantum numbers
through the $K^+p\to \pi^+K^+n$
reaction}
\author{T.~Hyodo$^a$\footnote{E-mail :
hyodo@rcnp.osaka-u.ac.jp}, A.~Hosaka$^a$ and E.~Oset$^b$}
\date{\today}
\maketitle

\begin{center}
$^a$\textit{Research Center for Nuclear Physics (RCNP),
Ibaraki, Osaka 567-0047, Japan} \\
$^b$\textit{Departmento de F\'isica Te\'orica and IFIC,
Centro Mixto Universidad de Valencia-CSIC,
Institutos de Investigaci\'on de Paterna, Aptd. 22085, 46071
Valencia, Spain}
\end{center}

\vspace{0.4cm}

\begin{abstract}
    We study the $K^+p\to \pi^+K^+n$ reaction with some kinematics
    suited to the production of the $\Theta^+$ resonance recently
    observed.
    We show that, independently of the quantum numbers of the
    $\Theta^+$, a resonance signal is always observed in the $K^+$
    forward direction.
    In addition, we also show how a combined consideration of the strength
    at the peak, and the angular dependence of polarization observables can help
    determine the $\Theta^+$ quantum numbers using 
    the present reaction.
\end{abstract}


A recent experiment by LEPS collaboration
at SPring-8/Osaka~\cite{Nakano:2003qx}
has found a clear signal for an $S=+1$ positive charge resonance
around 1540 MeV.
The finding, also confirmed by DIANA at
ITEP~\cite{Barmin:2003vv},
CLAS at Jefferson Lab.~\cite{Stepanyan:2003qr}
and SAPHIR at ELSA~\cite{Barth:2003es},
might correspond to the exotic state predicted by Diakonov {\em et~al.}
in Ref.~\cite{Diakonov:1997mm}.

Since the mass appears almost precisely at the value of the theoretical
prediction of Ref.~\cite{Diakonov:1997mm}
using the chiral quark soliton model,
the observed state may be identified with the
$\Theta$-baryon~\footnote{
In Ref.~\cite{Diakonov:1997mm}, this state was denoted as $Z^+$.  
However, as a standard
notation of baryon resonances, $\Theta^+$ has been recently proposed
by Diakonov.}
($(I,J^P) =(0, 1/2^+)$)
of five quarks ($uudd\bar s$), although spin, parity and isospin are not
yet determined experimentally.
Theoretically, the parity of $\Theta^+$ is interesting and 
important.
In the naive quark model, all quarks can be put in the lowest $1/2^+$
orbit.
Since $\bar s$ carries negative parity, the $\Theta (uudd \bar s)$ in
this naive picture would have negative parity,
in contrast with the prediction of the chiral quark soliton 
model~\cite{Diakonov:1997mm}.
There are several theoretical studies for this resonance
~\cite{Capstick:2003iq,Stancu:2003if,
Karliner:2003sy,hosaka,Jaffe:2003sg}
and recently lattice simulation has been also 
carried out~\cite{Csikor:2003ng}.

The experiment at SPring-8 was performed with the reaction
$\gamma n\to K^-K^+n$ with the $K^+n$ mass distribution showing the
peak of the resonance ($\Theta^+$).
The target neutron was in a $^{12}C$ nucleus,
but a test was carefully carried out in order to select
neutrons from the Fermi sea with small longitudinal momentum
which accurately gave the invariant mass 
of the $K^+n$ system by evaluating ($p_{\gamma}+p_n-p_{K^-}$)
with the assumption that the
initial neutron is at rest.
The test is convincing by selecting the final $K^+$ events
moving predominantly in the forward direction.
However, a dispersion of the invariant mass of around 10 MeV from the Fermi
motion is unavoidable.
In the experiment at Jefferson Lab, the reaction $\gamma d\to
K^+K^-pn$ was performed.
The measurement of the charged final particles
determines completely the kinematics and allows one to reconstruct
the $K^+n$ invariant mass where the $\Theta^+$ is seen.
Yet, elimination of the $K^+K^-$ resonant peak of the $\phi$ and
the $\Lambda(1520)$ for $K^-p$ events is needed to extract the
signal.
Also the dynamical mechanisms of photo-induced reactions in this energy
region are generally very complicated.

In such a situation,
alternative reactions based on known elementary processes are most welcome 
in order to increase our knowledge of the resonance and help
determine properties like spin, isospin and parity.
Some hadron-induced and photon-induced reactions 
are studied theoretically
in Refs.~\cite{Liu:2003rh,Nam:2003uf,Liu:2003zi}.
We present one particularly suited reaction with the process
\begin{equation}
    K^+p\to \pi^+K^+n \ .
    \label{eq:reaction}
\end{equation}
Characteristic features of this reaction are:
\begin{enumerate}

    \item The $K^+n$ invariant mass can be precisely determined by
measuring the $\pi^+$ momentum alone.

\item  It is expected that the initial state of the reaction is rather
simple, since we do not know of a $K^+p$ resonance that could 
complicate the initial system.

\item  By choosing small momenta of the $\pi^+$, we shall be also far
away from the $\Delta^+$ resonance and hence the $K^+n(\Theta^+)$
resonance signal can be more clearly seen.

\end{enumerate}

\begin{figure}[tbp]
    \centering
    \includegraphics[width=11cm,clip]{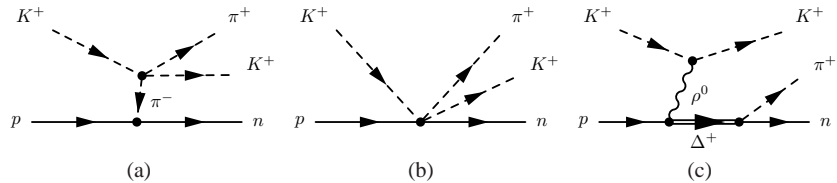}
    \caption{\label{fig:1}
    Feynman diagrams of the reaction $K^+p\to \pi^+K^+n$
    in the model of Ref.~\cite{Oset:1996ns}.}
\end{figure}

\begin{figure}[tbp]
    \centering
    \includegraphics[width=11cm,clip]{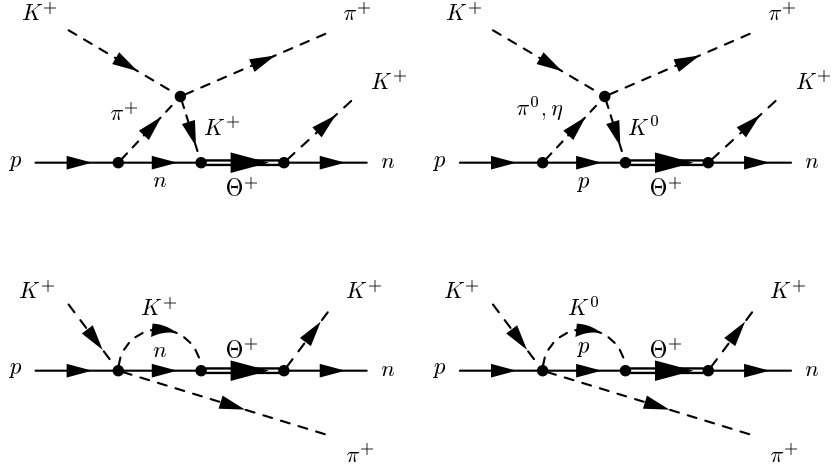}
    \caption{\label{fig:2}
    Feynman diagrams of the reaction $K^+p\to \pi^+K^+n$
    with the $\Theta^+$ resonance.}
\end{figure}

A successful model for the reaction~\eqref{eq:reaction} was
considered in
Ref.~\cite{Oset:1996ns}, consisting of the mechanisms depicted in
terms of Feynman diagrams in Fig.~\ref{fig:1}.
The term (a) (pion pole) and (b) (contact term), which are easily
obtained from the chiral Lagrangians involving
meson-meson~\cite{Gasser:1985gg,Meissner:1993ah}
and meson-baryon interaction~\cite{Pich:1995bw,Ecker:1995gg,Bernard:1995dp}
are spin flip terms (proportional to $\bm{\sigma}$), while the $\rho$
exchange term (diagram (c)) contains both a spin flip and a non spin 
flip part. Having an amplitude proportional to $\bm{\sigma}$ is
important in the present context in order to have a test of the
parity of the resonance.
Hence we choose a situation, with the final pion momentum 
$\bm{p}_{\pi^+}$ small compared to 
the momentum of the initial kaon, such that the diagram (c), which 
contains the $\bm{S}\cdot \bm{p}_{\pi^+}$ operator can be safely
neglected.
The terms of Fig.~\ref{fig:1} (a) and (b) will provide the bulk for
this reaction.
If there is a resonant state for $K^+n$ then this 
will be seen in the final state interaction of this system.
This means that in addition to the diagrams (a) and (b) 
of Fig.~\ref{fig:1},
we shall have those in Fig.~\ref{fig:2}.
If the resonance is an $s$-wave $K^+n$ resonance then $J^P=1/2^-$.
If it is a $p$-wave resonance, we can have $J^P=1/2^+, 3/2^+$.
We shall study all these three possibilities,
but couplings of higher partial waves are not considered here.

\begin{figure}[tbp]
    \centering
    \includegraphics[width=6cm,clip]{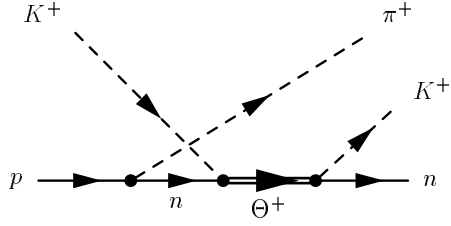}
    \caption{\label{fig:new}
    Feynman diagrams of the reaction $K^+p\to \pi^+K^+n$
    with the $\Theta^+$ resonance.}
\end{figure}

The restriction to have small pion momenta eliminates also other 
possible resonant contributions like the one in the diagram 
of Fig.~\ref{fig:new},
which involves the $\bm{\sigma}\cdot \bm{p}_{\pi^+}$ coupling.
One could in principle have negative parity resonances instead of the
intermediate neutron, which involve  an s-wave $\pi N N^*$ coupling, but the
lightest one would be the $N^*(1535)$ which would be very much off shell,
rendering the contribution of the diagram negligible. Note also that the
consideration of $N^*$ resonances instead of the intermediate $n$, 
$p$ states in
Fig~\ref{fig:2}
is not usually done in chiral dynamics calculations since it is accounted
for by means of counterterms, or subtraction constants in the loop function
which can be equally taken into account by the choice of the cut off 
\cite{Oller:2000fj}.

The resonance contribution to
the $K^+n(K^0p) \to K^+n$ amplitudes are given by
\begin{equation}
    \begin{split}        
	t^{(s)}_{K^+n(K^0p)\to K^+n}
	&=\frac{(\pm)g_{K^+n}^2}{M_I-M_R+i\Gamma/2} \ , \\
	t^{(p,1/2)}_{K^+n(K^0p)\to K^+n}
	&=\frac{(\pm)\bar{g}_{K^+n}^2
	(\bm{\sigma}\cdot \bm{q}^{\prime})
	(\bm{\sigma}\cdot \bm{q})}{M_I-M_R+i\Gamma/2} \ , \\
	t^{(p,3/2)}_{K^+n(K^0p)\to K^+n}
	&=\frac{(\pm)\tilde{g}_{K^+n}^2
	(\bm{S}\cdot \bm{q}^{\prime})
	(\bm{S}^{\dag}\cdot \bm{q})}{M_I-M_R+i\Gamma/2} \ ,
    \end{split}
    \label{eq:SPresonance}
\end{equation}
for $s$-wave and $p$-wave with $J^P=1/2^+,3/2^+$,
where $M_I$ is the energy of the $K^+n$
system in the center of mass frame, $M_R$ and $\Gamma$ the mass and
width of the resonance $(1540\, \text{MeV},20\, \text{MeV})$,
and  $q$ and $q^{\prime}$ the momenta of the initial
and final kaon, respectively.
$\bm{S}^{\dag}$ is the spin transition operator from spin 
$1/2$ to $3/2$ states.
The signs $+$ and $-$ stand for the $K^0p\to K^+n$ amplitude in the case
one has $I=1$ or $I=0$ for the $\Theta^+$ resonance.
We are implicitly assuming isospin conservation and hence
we do not consider the possibility of $I=2$,
suggested in Ref.~\cite{Capstick:2003iq},
which implies isospin violation in its decay to $KN$.
The values of $g$, $\bar{g}$ and $\tilde{g}$
can be obtained from the $\Theta^+$
width
\begin{equation}
        g_{K^+n}^2
	=\frac{\pi M_R\Gamma}{Mq} \ , \quad
	\bar{g}_{K^+n}^2
	=\frac{\pi M_R\Gamma}{Mq^3} \ , \quad
	\tilde{g}_{K^+n}^2
	=\frac{3\pi M_R\Gamma}{Mq^3} \ .
    \label{eq:couplings}
\end{equation}

A straightforward evaluation of the meson pole and contact terms (see also
Ref.~\cite{Meissner:1995vp}) leads to the $K^+n\to \pi^+KN$
amplitudes
\begin{equation}
    -it_i
    =(a_i+b_i\bm{k}_{in}\cdot \bm{q}^{\prime}+c_i)\bm{\sigma}
    \cdot\bm{k}_{in}
    +(-a_i-b_i\bm{k}_{in}\cdot \bm{q}^{\prime}+d_i)\bm{\sigma}
    \cdot\bm{q}^{\prime} \ ,
    \label{eq:t1amp}
\end{equation}
where $i=1,2$ stands for the final state $K^+n, K^0p$ respectively
and $k_{in}$ and $q^{\prime}$ are the initial and final $K^+$ momenta.
The coefficients $a_i$ and $b_i$ are from meson exchange terms, and
$c_i$ and $d_i$ from contact terms.
They are given by
\begin{align}
    a_1
    =&-\frac{1}{3f^2}(m_K^2-2m_{\pi}\omega(q)-\omega(k_{in})\omega(q)
    -\omega(k_{in})m_{\pi}) \nonumber\\ 
    &\cdot
    \frac{\sqrt{2}(D+F)}{2f}\frac{1}{p_{ex}^2-m_{\pi}^2} \ ,
    \label{eq:a1} \\
    a_2
    =&a_{2,\pi}+a_{2,\eta} \ ,\\
    a_{2,\pi}
    =&-\frac{1}{\sqrt{2}f^2}
    m_{\pi}(\omega(k_{in})+\omega(q)) \cdot
    \frac{D+F}{2f}\frac{1}{p_{ex}^2-m_{\pi}^2} \ , \\
    a_{2,\eta}
    =&\frac{1}{\sqrt{6}f^2}(
    2\omega(k_{in})\omega(q)+m_{\pi}\omega(q)-m_{\pi}\omega(k_{in})
    -\frac{2}{3}m_{K}^2+\frac{2}{3}m_{\pi}^2) \nonumber\\
    & \cdot
    \frac{3F-D}{2\sqrt{3}f}\frac{1}{p_{ex}^2-m_{\eta}^2} \ , \\
    b_1
    =&-\frac{1}{3f^2}\cdot \frac{\sqrt{2}(D+F)}{2f}
    \frac{1}{p_{ex}^2-m_{\pi}^2} \ ,\\
    b_2
    =& b_{2,\pi}+b_{2,\eta} \ , \\
    b_{2,\pi}
    =&0 \ ,\\
    b_{2,\eta}
    =&-\frac{1}{\sqrt{6}f^2}\cdot \frac{3F-D}{\sqrt{3}f}
    \frac{1}{p_{ex}^2-m_{\eta}^2} \ ,\\
    c_1
    =& \frac{\sqrt{2}(D+F)}{12f^3} \ ,\\
    c_2
    =&- \frac{\sqrt{2}D}{12f^3} \ ,\\
    d_1
    =&\frac{\sqrt{2}(D+F)}{24f^3} \ , \\
    d_2
    =&-\frac{D+3F}{2}\frac{\sqrt{2}}{12f^3} \ , \label{eq:d2}
\end{align}
where 
$p_{ex}$ is the momentum of the meson exchanged in the meson pole term.
The three momentum of the final pion is already neglected in these
formulae.

Now let us turn to the resonance diagrams of Fig.~\ref{fig:2}
containing a loop integral, which is initiated by the tree diagrams
of Figs.~\ref{fig:1} (a) and (b).
The momentum of $K^+$ is now an internal variable $\bm{q}$.
In performing the loop integral,
the fact that $\bm{k}_{in}$ is reasonably larger than $\bm{q}$,
allows us to make an angle average of the meson propagator
which simplifies the integrals.
Furthermore, as shown in Ref.~\cite{Hyodo:2003jw}
and also found in the meson baryon
scattering processes~\cite{Oset:1998it}, the amplitude $K^+p\to
\pi^+K^+n$ factorizes inside the loops with its on-shell value, which
means in Eqs.~\eqref{eq:a1}-\eqref{eq:d2}  one takes
\begin{equation}
    \omega(q)=\frac{M_I^2+m_K^2-M_N^2}{2M_I}
    \nonumber
\end{equation}
Note that since we have chosen the $\pi^+$ momentum small,
the $K^+n$ final state is also approximately in its center of mass
frame and we assume this kinematics in the variables.

When taking into account $KN$ scattering through the $\Theta^+$
resonance, as depicted in Fig.~\ref{fig:2}, the $K^+p\to \pi^+K^+n$
amplitude is given by
\begin{equation}
    -i\tilde{t}=-it_1-i\tilde{t}_1-i\tilde{t}_2
    \label{eq:total}
\end{equation}
where $\tilde{t}_1$ and $\tilde{t}_2$ account for the scattering
terms with intermediate $K^+n$ and $K^0p$, respectively.
They are given by
\begin{equation}
\begin{split}
    -i\tilde{t}^{(s)}_i
    =&\frac{g_{K^+n}^2}{M_I-M_R+i\Gamma/2}
    \left\{G(M_I)(a_i+c_i)
    -\frac{1}{3}\bar{G}(M_I)b_i\right\}
    \bm{\sigma}\cdot\bm{k}_{in}S_I(i) \ ,\\
    -i\tilde{t}^{(p,1/2)}_i
    =&\frac{\bar{g}_{K^+n}^2}{M_I-M_R+i\Gamma/2}
    \bar{G}(M_I)
    \left\{\frac{1}{3}b_i\bm{k}_{in}^2-a_i+d_i
    \right\}
    \bm{\sigma}\cdot\bm{q}^{\prime} S_I(i) \ , \\
    -i\tilde{t}^{(p,3/2)}_i
    =&\frac{\tilde{g}_{K^+n}^2}{M_I-M_R+i\Gamma/2}
    \bar{G}(M_I)
    \frac{1}{3}b_i
    \left\{
    (\bm{k}_{in}\cdot\bm{q}^{\prime})
    (\bm{\sigma}\cdot\bm{k}_{in})
    -\frac{1}{3}\bm{k}_{in}^2\bm{\sigma}\cdot\bm{q}^{\prime}
    \right\}S_I(i) \ , 
\end{split}
    \label{eq:tilde}
\end{equation}
for $s$- and $p$-wave, and $i=1,2$ for $K^+n$ and $K^0p$
respectively.
$S_I(i)$ gives the sign for the $K^+n$ and $K^0p$ components in $I=0$
and $1$. Thus $S_0(1)=1$, $S_1(1)=1$, $S_0(2)=-1$ and $S_1(2)=1$.
The function $G(M_I)$ is the loop function of a meson and  a baryon
propagator
\begin{equation}
    G(M_I)
    =\int \frac{d^3q}{(2\pi)^3}
    \frac{1}{2\omega(q)}\frac{M}{E(q)}
    \frac{1}{M_I-\omega(q)-E(q)+i\epsilon} \ ,
    \label{eq:Gfn}
\end{equation}
regularized in Ref.~\cite{Oset:1998it} with a three momentum
cut off $q_{max}=630$ MeV/c to reproduce the data of $\bar{K}N$
scattering.
We use here the same function.
Similarly, $\bar{G}(M_I)$ is the same integral but with a factor
$\bm{q}^2$ in the numerator of Eq.~\eqref{eq:Gfn}.

For completeness, we include a recoil factor 
in all terms to account for $\mathcal{O}(p/M)$ relativistic
corrections for the
$\gamma^{\mu}\gamma_5\partial_{\mu}$ BBM vertex,
which is given by
\begin{equation}
    F_i=\left(1-\frac{p_{ex}^{0(i)}}{2M_p}\right)
    \label{eq:recoil} \ .
\end{equation}
In addition, we also consider the strong form factor of the $MMB$
vertex for which we take a standard monopole and static
form factor
\begin{equation}
    F_f(\bm{p})=\frac{\Lambda^2-m_{\pi}^2}{\Lambda^2+\bm{p}^2}
    \label{eq:FF}
\end{equation}
with $\Lambda\sim 900$ MeV.
This form factor is applied both to the meson pole and contact terms 
to preserve the subtle cancellation of off shell terms shown in
Ref.~\cite{Hyodo:2003jw}. Inside the loops, for the reasons exposed above,
 the product of the form factor and
propagator is replaced by its angle averaged expression which simplifies the
formulae. This is implicit in the $a_i,b_i$ coefficients of 
Eqs.~\eqref{eq:tilde}.

We take an initial three momenta of $K^+$ in the Laboratory frame
$k_{in}(Lab)=850 \,\,\text{MeV}/c$ 
($\sqrt{s}=1722$ MeV), which allows us to
span $K^+n$ invariant masses up to $M_I=1580$ MeV, thus covering
the peak of the $\Theta^+$, and still is small enough to have negligible
$\pi^+$ momenta with respect to the one of the incoming $K^+$.
The double differential cross section is given by ($\cos \theta$ is
the angle between $\bm{k}_{in}$ and $\bm{q}^{\prime}$
\begin{equation}
\begin{split}
    \frac{d^2\sigma}{dM_Id\cos\theta}
    =& \frac{1}{(2\pi)^3}\frac{1}{8s}
    \frac{M^2}{\lambda^{1/2}(s,M^2,m_K^2)}
    \frac{1}{M_I} \\
    &\times\bar{\Sigma}\Sigma|t|^2
    \lambda^{1/2}(s,M_I^2,m_{\pi}^2)
    \lambda^{1/2}(M_I^2,M^2,m_{K}^2) \ .
\end{split}
    \label{eq:dcross}
\end{equation}
where $\lambda(x,y,z)$ is the K\"allen function defined by
$\lambda(x,y,z)=x^2+y^2+z^2-2xy-2yz-2zx$.
We show below the results for the different options of isospin,
spin and parity of the $\Theta^+$.

\begin{figure}[tbp]
    \centering
    \includegraphics[width=11cm,clip]{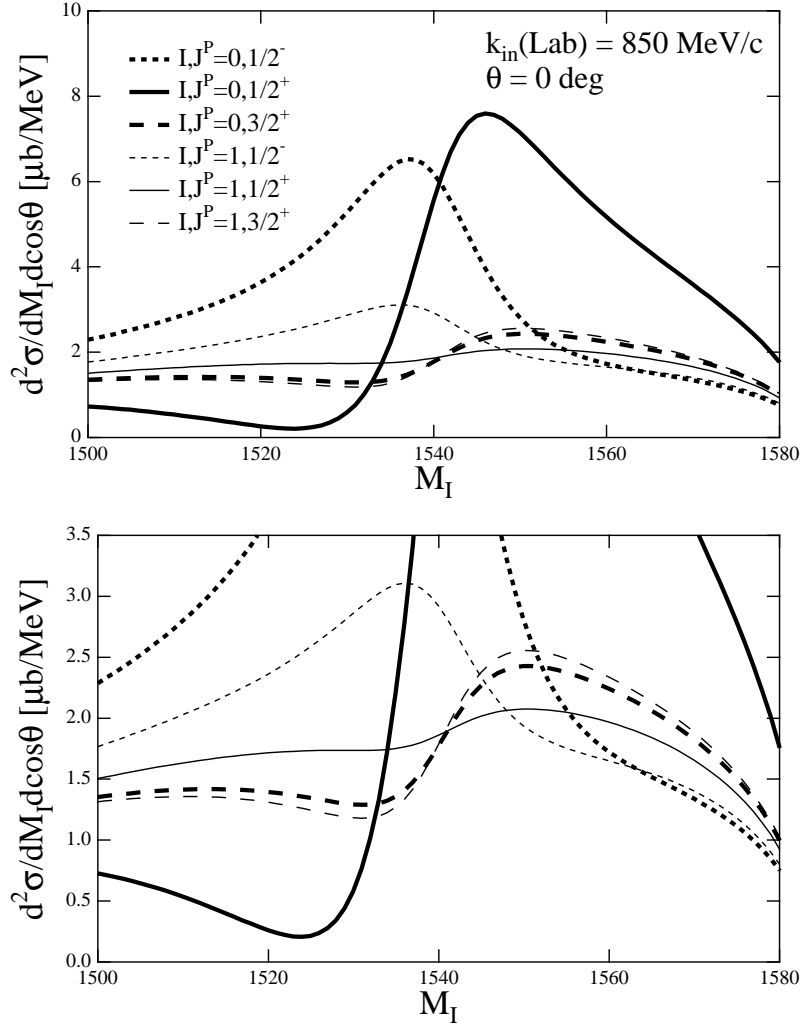}
    \caption{\label{fig:3}
    The double differential cross sections 
    $d^2\sigma/dM_Id\cos\theta$ with $\theta=0$ (forward direction)
    for $I=0,1$ and $J^P=1/2^-,1/2^+,3/2^+$.
    Below, detail of the lower part of the upper figure of the panel.}
\end{figure}

In Fig.~\ref{fig:3}, we show the invariant mass distribution
$d^2\sigma/dM_Id\cos\theta$ in the $K^+$
forward direction ($\theta=0)$.
Here we see that, independently of the quantum numbers of
$\Theta^+$, a resonance signal is always observed.
The signals for the resonance are quite clear for the case
of $I,J^P=0,1/2^+$ (these would be the quantum numbers predicted 
in Ref.~\cite{Diakonov:1997mm}) and  $I,J^P=0,1/2^-$, while in the other
cases the signal is weaker and the background more important,
particularly for the case of $I,J^P=0,1/2^+$. 
With estimated uncertainties in the theory of the order of 
20-30 percent, from the approximations done, 
the strength of the peak at the resonance could already serve
to discriminate among the several possibilities. 

We have used a $\Theta ^+$ width of 20 MeV,
but experimentally it could be
smaller since the experimental widths observed are mostly
coming from the experimental resolution~\cite{Nussinov:2003ex,
Arndt:2003xz}. 
It is easy to see how this would change our results.
By looking at Eqs.~\eqref{eq:tilde} at the peak of the
resonance distribution and considering the couplings
obtained in Eq.~\eqref{eq:couplings} 
we see that the strength at the peak is independent of $\Gamma$.
Only the width of the calculated distributions would be smaller
for smaller  $\Gamma$.

\begin{figure}[tbp]
    \centering
    \includegraphics[width=11cm,clip]{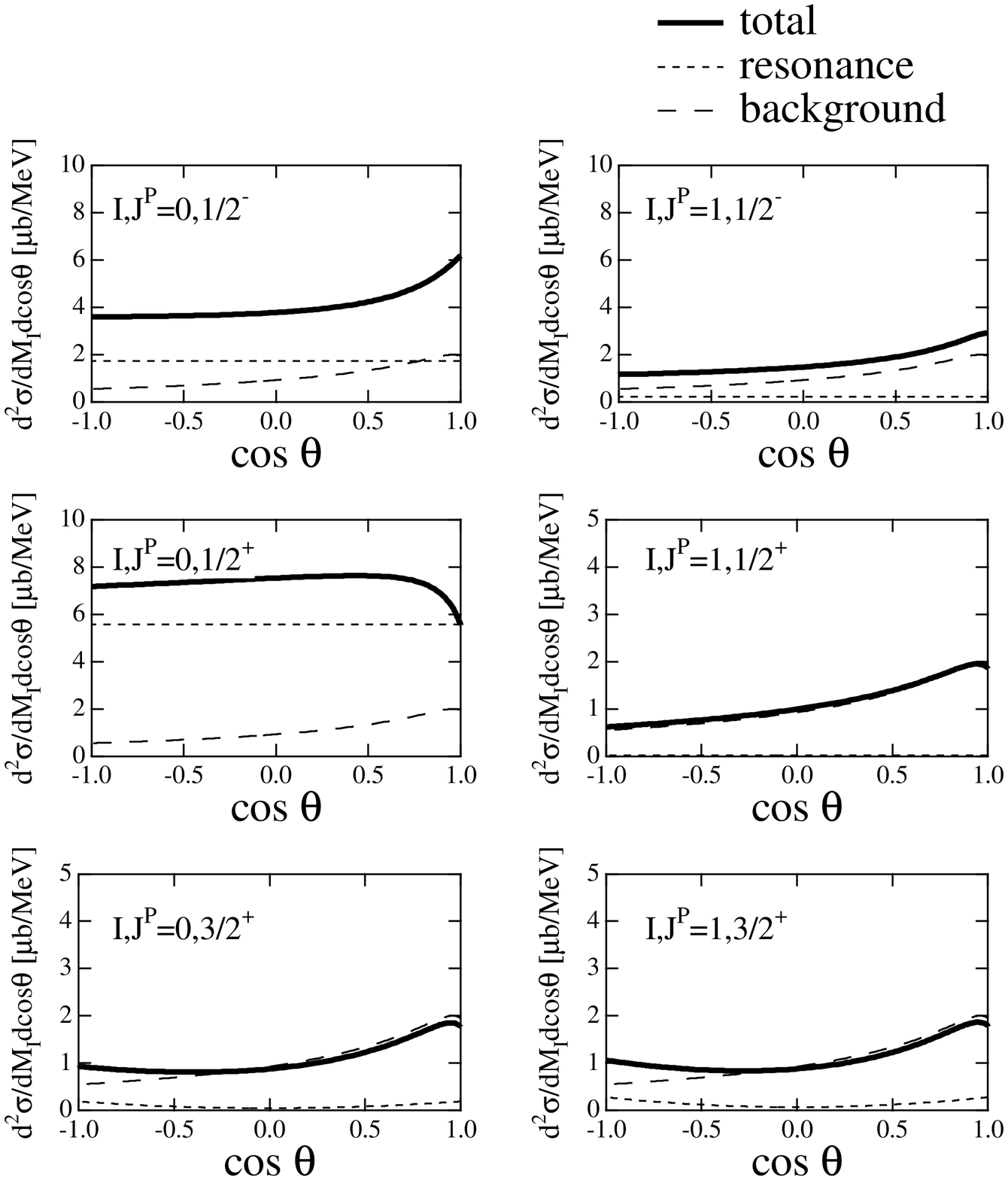}
    \caption{\label{fig:4}
    Angular dependence of the double differential cross sections
    $d^2\sigma/dM_Id\cos\theta$ with $M_I=1540$ MeV at the resonance peak
    for $I=0,1$ and $J^P=1/2^-,1/2^+,3/2^+$.}
\end{figure}

The angular dependence is shown in Fig.~\ref{fig:4} 
for a value of $M_I=1540$ MeV. 
What we observe there is that the angular dependence is 
rather weak in all cases. 
The background has a weak angular dependence and the resonance signal
for this unpolarized cross section has only
angular dependence for the case of spin 3/2, where it goes as 
$(3 \cos^2\theta +1)$, but in this case the resonance contribution 
is much smaller than the background and this angular dependence
is not very visible. 
The different inflexions of the cross section at $\theta =0$
are probably too small to be discriminated in an experiment,
hence the conclusion is that this unpolarized observable does not shed
any further light on the quantum numbers.

Let us now see what can one learn with resorting to polarization
measurements. 
Eqs.~\eqref{eq:tilde} account for the resonance contribution
to the process.
The interesting finding there is that if the $\Theta^+$ 
couples to $K^+n$ in $s$-wave (hence negative parity) 
the amplitude goes as $\bm{\sigma}\cdot \bm{k}_{in}$ 
while if it couples in $p$-wave it has a term
$\bm{\sigma}\cdot \bm{q}^{\prime}$.
Hence, a possible polarization test to determine which one of the
couplings the resonance chooses is to measure the cross section for
initial proton polarization $1/2$ in the direction $z$ $(\bm{k}_{in})$
and final neutron polarization $-1/2$  (the experiment
can be equally done with $K^0p$ in the final state, which makes the
nucleon detection easier).
In this spin flip amplitude $\langle -1/2 |t|+1/2\rangle$, the 
$\bm{\sigma}\cdot \bm{k}_{in}$ term vanishes.
With this test the resonance signal disappears for the $s$-wave
case, while the $\bm{\sigma}\cdot \bm{q}^{\prime}$ operator of the
$p$-wave case would have a finite matrix element 
proportional to $q^{\prime}\sin \theta$.
This means, away from the forward direction of the final kaon, the
appearance of a resonant peak in the cross section would indicate a
$p$-wave coupling and hence a positive parity resonance.

\begin{figure}[tbp]
    \centering
    \includegraphics[width=11cm,clip]{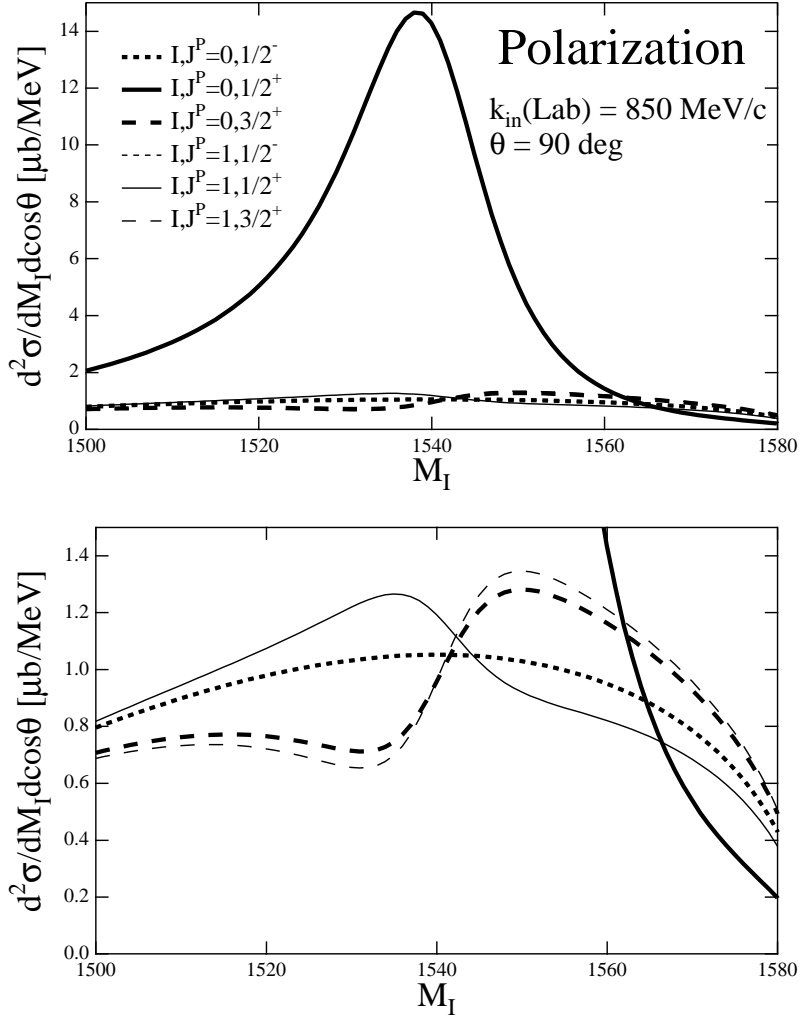}
    \caption{\label{fig:5}
    The double differential cross sections
    $d^2\sigma/dM_Id\cos\theta$ with $\theta=0$ (forward direction)
    for $I=0,1$ and $J^P=1/2^-,1/2^+,3/2^+$.
    Below, detail of the lower part of the upper figure of the panel.
    In these figures, thick and thin
    short-dashed lines ($J^P = 1/2^-$) coincide, 
    since there is only background contribution.
    In the upper panel, the almost identical results
    of $J^P = 3/2^+$ drawn by the two long-dashed lines
    also coincide.}
\end{figure}

In Fig.~\ref{fig:5} we show the results for the polarized cross section
measured at 90 degrees as a function of the invariant mass.
The two cases with s-wave do not show any resonant shape
since only the background contributes.
All the other cross sections are quite reduced to the point
that the only sizeable resonant peak comes from the $I,J^P=0,1/2^+$ case.
A clear experimental signal of the resonance in this observable
would unequivocally indicate the quantum numbers as  $I,J^P=0,1/2^+$.

\begin{figure}[tbp]
    \centering
    \includegraphics[width=11cm,clip]{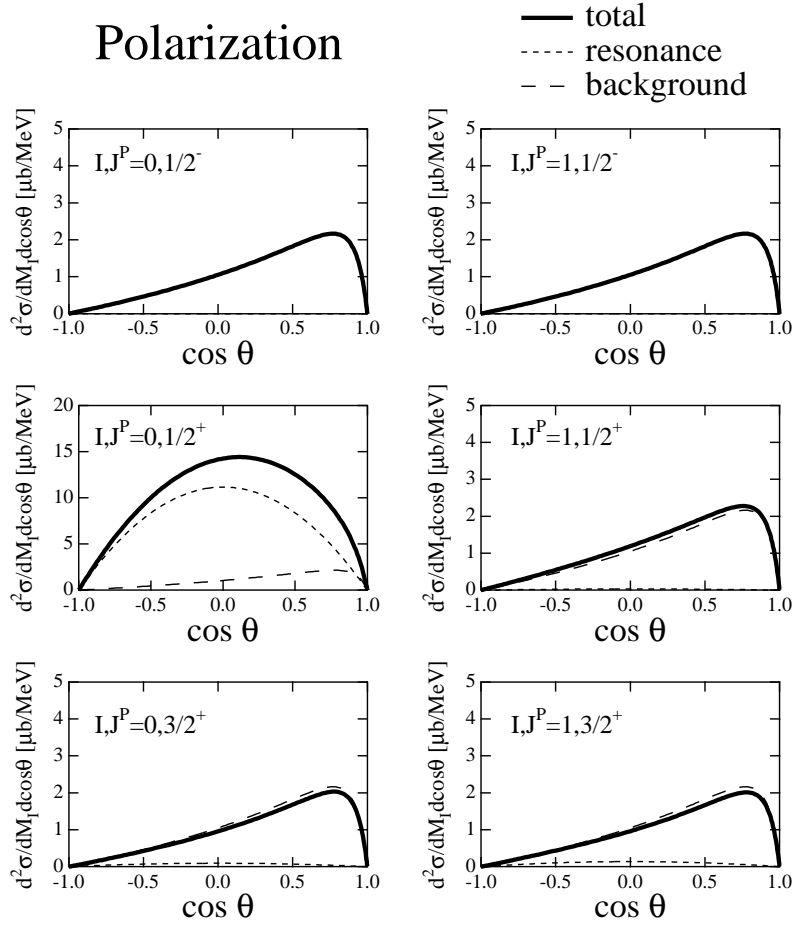}
    \caption{\label{fig:6}
    Angular dependence of the double differential cross sections
    of polarized amplitude with $M_I=1540$ MeV at the resonance peak
    for $I=0,1$ and $J^P=1/2^-,1/2^+,3/2^+$.}
\end{figure}

Finally, in Fig.~\ref{fig:6} we show the angular dependence 
of the polarized cross section for a fixed value of 
the invariant mass of 1540 MeV.
The angular distributions look all of them similar,
as a consequence of the weakness or disappearance of the resonance 
contribution, with a peak around 35 degrees, 
except for the case of $I,J^P=0,1/2^+$, 
where the peak is found around 80 degrees 
and has a much larger size than in the other cases.

\begin{figure}[tbp]
    \centering
    \includegraphics[width=11cm,clip]{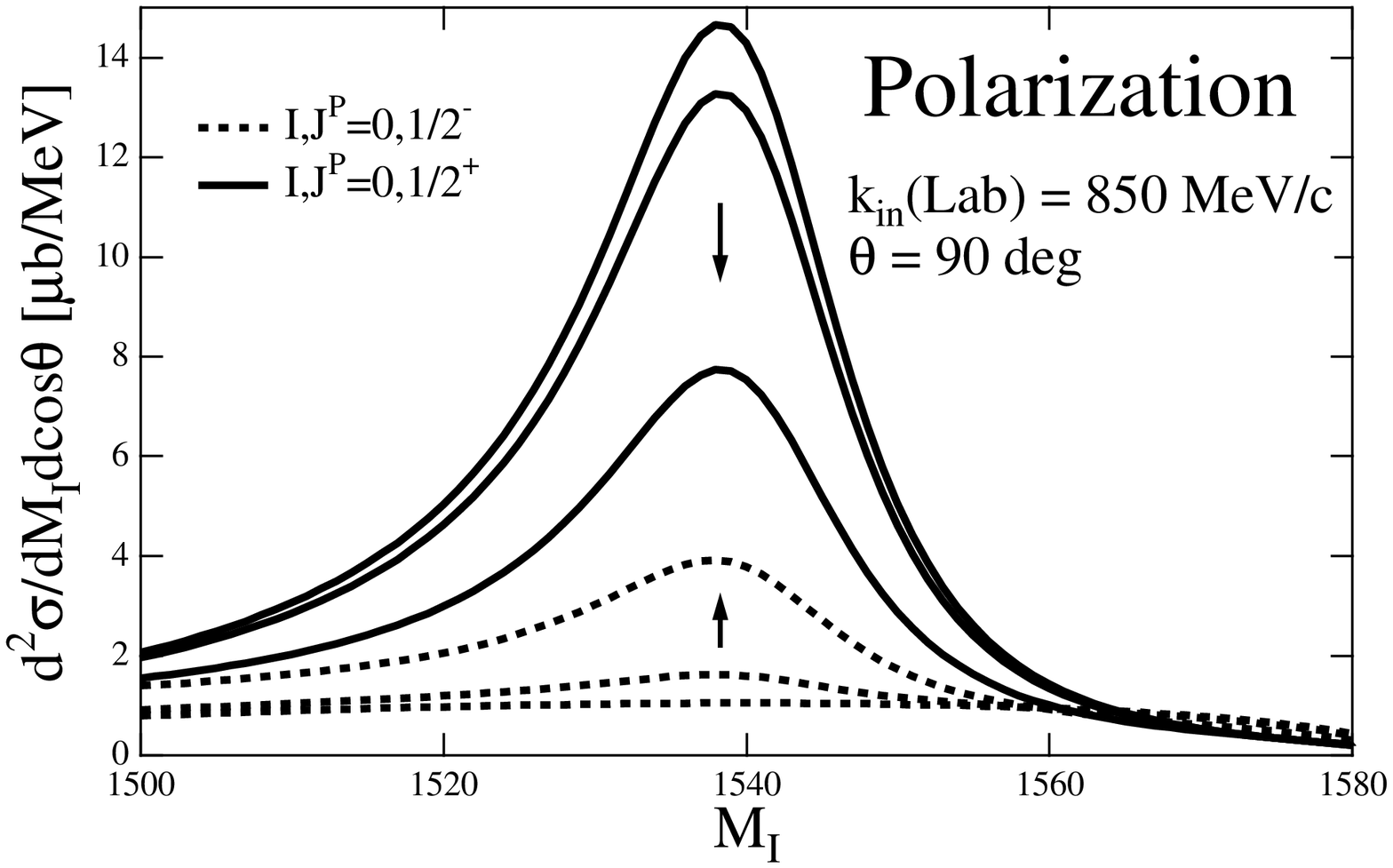}
    \caption{\label{fig:new2}
    Effects of incomplete polarization on the double differential
    cross sections at $\theta = 90$ degrees.
    Three solid and dashed lines are for
    100 \%, 80 \% and 0 \% (along the direction of an arrow)
    polarization for $J^P= 1/2^{\pm}$ and $I = 0$.}
\end{figure}

Since 100 \% polarization can not be achieved
in actual experiments, we have computed cross
sections for the case of incomplete polarization.
We have then found that for a typical polarization
rate of about 80 \%~\footnote{
We define the polarization rate by $(N_+ - N_-)/(N_+ + N_-)$.},
the previous results shown
in Figs. 6 and 7 do not change much.
For instance, as shown in Fig.~\ref{fig:new2},
the cross section decreases
about 10 \% for $I, J^P = 0, 1/2^+$.
For $I, J^P = 0, 1/2^-$, the peak value at around
$M_I \sim 1540$ MeV increases, since
there is no resonance contribution for the case of 100 \%
polarization.
However, the absolute value is small as compared to
the one of $J^P = 1/2^+$.
For the totally unpolarized case, the $1/2^+$
cross section reduces to about half of the polarized
one, while the $1/2^-$ cross section shows a sizable peak.
The tendency for other cross sections such as angular
dependence is also similar.

In summary, we have shown here an elementary reaction, $K^+p\to
\pi^+K^+n$, where, based on the present knowledge of the  $\Theta^+$
resonance, we can make predictions for $\Theta^+$ 
production cross sections.
We see that, independently of the $\Theta^+$ quantum numbers,
a resonance signal is always seen in the forward direction of the
final $K^+$. 
The strength at the peak could serve to discriminate among several cases. 
Further measurements of a polarized cross section could serve to further
eliminate other possibilities. 
In particular, a strong signal seen at 90 degrees
for the polarized cross section
would clearly indicate that the quantum numbers
of the resonance are those predicted in Ref.~\cite{Diakonov:1997mm}.

The reaction suggested here can be easily performed and, in
particular, a small change in the set up of the experiment at ITEP used
to detect the $\Theta^+$ could be made to perform the reaction.
The determination of the quantum numbers of the $\Theta^+$ is an
essential step to further investigate its nature.
The implementation of the present reaction would represent
a step forward in this direction.

\section*{Acknowledgments}
This work is supported by the Japan-Europe (Spain) Research
Cooperation Program of Japan Society for the Promotion of Science
(JSPS) and Spanish Council for Scientific Research (CSIC), which
enabled E. O. to visit RCNP, Osaka and T. H. and A. H. to visit
IFIC, Valencia.
This work is also supported in part  by DGICYT
projects BFM2000-1326,
and the EU network EURIDICE contract
HPRN-CT-2002-00311.

%

\end{document}